# Electronic Structures and Bonding of Oxygen on Plutonium Layers

M. N. Huda and A. K. Ray*
*P. O. Box 19059, Department of Physics, The University of Texas at Arlington
Arlington, Texas-76013*

Oxygen adsorptions on $\delta$-Pu (100) and (111) surfaces have been studied at both non-spin-polarized and spin-polarized levels using the generalized gradient approximation of density functional theory (GGA-DFT) with Perdew and Wang (PW) functionals. The center position of the (100) surface is found to be the most favorable site with chemisorption energies of 7.386eV and 7.080eV at the two levels of theory. The distances of the oxygen adatom from the Pu surface are found to be 0.92A and 1.02A, respectively. For the (111) surface non-spin-polarized calculations, the center position is also the preferred site with a chemisorption energy of 7.070eV and the distance of the adatom being 1.31A, but for spin-polarized calculations the bridge and the center sites are found to be basically degenerate, the difference in chemisorption energies being only 0.021eV. In general, due to the adsorption of oxygen, plutonium 5f orbitals are pushed further below the Fermi energy, compared to the bare plutonium layers. The work function, in general, increases due to oxygen adsorption on plutonium surfaces.

## A. Introduction

Considerable theoretical efforts have been devoted in recent years to studying the electronic and geometric structures and related properties of surfaces to high accuracy. One of the many motivations for this burgeoning effort has been a desire to understand the detailed mechanisms that lead to surface corrosion in the presence of environmental gases; a problem that is not only scientifically and technologically challenging but also environmentally important. Such efforts are particularly important for systems like the actinides for which experimental work is relatively difficult to perform due to material problems and toxicity. As is known, the actinides are characterized by a gradual filling of the 5f-electron shell with the degree of localization increasing with the atomic number Z along the last series of the periodic table. The open shell of the 5f electrons determines the magnetic and solid state properties of the actinide elements and their compounds and understanding the quantum mechanics of the 5f electrons is the defining issue in the physics and chemistry of the actinide elements. These elements are also characterized by the increasing prominence of relativistic effects and their studies can, in fact, help us understand the role of relativity throughout the periodic table. Narrower 5*f* bands near the Fermi level, compared to 4*d* and 5*d* bands in transition elements, is believed to be responsible for the exotic structure of actinides at ambient condition.[1] The 5f orbitals have properties intermediate between those of localized 4f and delocalized 3d orbitals and as such, the actinides constitute the "missing link" between the d transition elements and the lanthanides.[2] Thus a proper and accurate understanding of the actinides will help us understand the behavior of the lanthanides and transition metals as well.

*email:akr@exchange.uta.edu

Among the actinides, plutonium is particularly interesting in two respects [3-6]. First, Pu has, at least, six stable allotropes between room temperature and melting at atmospheric pressure, indicating that the valence electrons can hybridize into a number of complex bonding arrangements. Second, plutonium represents the boundary between the light actinides, Th to Pu, characterized by itinerate 5f electron behavior, and the heavy actinides, Am and beyond, characterized by localized 5f electron behavior. In fact, the high temperature fcc δ-phase of plutonium exhibits properties that are intermediate between the properties expected for the light and heavy actinides. These unusual aspects of the bonding in bulk Pu are apt to be enhanced at a surface or in a ultra thin film of Pu adsorbed on a substrate, due to the reduced atomic coordination of a surface atom and the narrow bandwidth of surface states. For this reason, Pu surfaces and films and adsorptions on such may provide a valuable source of information about the bonding in Pu.

This work has concentrated on square and hexagonal Pu layers corresponding to the (100) and (111) surfaces of δ-Pu and adsorptions of oxygen adatoms on such surfaces, using the formalism of modern density functional theory. Although the monoclinic α-phase of Pu is more stable under ambient conditions, there are advantages to studying δ-like layers. First small amount of impurities can be used to stabilize δ-Pu at room temperature. Second, grazing-incidence photoemission studies combined with the calculations of Eriksson *et al.* [7] suggest the existence of a small moment δ-like surface on α-Pu. Our work on Pu monolayers has also indicated the possibility of such a surface [8]. Recently, high-purity ultrathin layers of Pu deposited on Mg were studied by X-ray photoelectron (XPS) and high-resolution valence band (UPS) spectroscopy by Gouder *et al* [9]. They found that the degree of delocalization of the 5f states depends in a very dramatic way on the layer thickness and the itinerant character of the 5f states is gradually lost with reduced thickness, suggesting that the thinner films are δ-like. The localised 5f states, which appear as a broad peak 1.6 eV below the Fermi level, were observed for one monolayer. At intermediate thickness, three narrow peaks appear close to the Fermi level and a comparative study of bulk α-Pu indicated a surface reorganization yielding more localized f-electrons at thermodynamic equilibrium. Finally, it may be possible to study 5f localization in Pu through adsorptions on carefully selected substrates in which case the adsorbed layers are more likely to be δ-like than α-like. We first comment on the published literature, followed by our results.

Experimental data [10] indicates that when Pu surface is exposed to molecular oxygen, oxygen is readily adsorbed by the metal surface. The oxygen molecule then dissociates into atomic oxygen, and combines with Pu to form a layer of oxide. Oxidation continues and the oxygen diffuses through the oxide layer reacting with more plutonium and producing more oxide at the oxide/metal interface, eventually reaching a steady state thickness. Using the film-linearized-muffin-tin-orbitals (FLMTO) method, Eriksson *et al.* [7] have studied the electronic structure of hydrogen and oxygen chemisorbed on Pu. The slab geometry was chosen to have the $CaF_2$ structure and the chemisorbed atoms were assumed to have fourfold-bridging positions at the surface. They found the surface behavior in $PuH_2$ and $PuO_2$ to be rather different compared to the surface behavior in pure metallic Pu. For metallic Pu, the 5f electrons are valence electrons and show only a small covalent like bonding contribution associated with small 5f to non-5f band hybridization. For the hydride and oxide, the Pu 5f electrons were well localized and treated as core electrons. Thus, the Pu valence behavior is dominated by the 6d electrons, giving rise to significant hybridization with ligand valence electrons and significant covalency. The energy gained when the H atoms chemisorb on the Pu surface was found to be 4.0 eV per atom. There are *no other theoretical studies* in the literature on oxygen adsorption on the Pu surface. In our previous hybrid density functional cluster study of the bulk and surface electronic structures of PuO [11], a large overlap between the Pu 5f bands and O 2p bands and a significant covalent nature in the chemical bonding were found. The highest occupied molecular orbital – lowest unoccupied molecular orbital (HOMO-LUMO) gaps and the density of states of the clusters supported the idea that PuO is a semiconductor. In follow-up studies of $PuO_2$ (110) surface and water adsorption on this surface, we have shown that the adsorption is dissociative and oxygen interaction is relatively strong [12]. In a recent study using the self-interaction corrected local spin density method, Petit *et al.* [13] reported the electronic structure of $PuO_{2\pm x}$. They found that in the stoichiometric $PuO_2$ compound, Pu occurs in the Pu (IV) oxidation state, corresponding to a localized $f^4$ shell. If oxygen is introduced onto the octahedral interstitial site, the nearby Pu atoms turn into Pu (V) ($f^3$) by transferring electron to the oxygen.

**B. Computational Details and Results**

As in our previous work [12], all computations reported here have been performed at the generalized gradient approximation (GGA) level [14] of density functional theory (DFT) [15] using the suite of programs DMol3 [16]. In DMol3, the physical wave function is expanded in accurate numerical basis set and fast convergent three-dimensional integration is used to calculate the matrix elements occurring in the Ritz variational method. For the oxygen adatom, a double numerical basis set with polarization functions (DNP) and real space cut-off of 4.5 Å was used. The sizes of these DNP basis set are comparable to the 6-31G** basis of Hehre *et al.* [17]. However, they are believed to be much more accurate than a Gaussian basis set of the same size [16]. For Pu, the outer sixteen electrons ($6s^2$ $6p^6$ $5f^6$ $7s^2$) are treated as valence electrons and the remaining seventy-eight electrons are treated as core. A hardness conserving semi-local pseudopotential, called density functional semi-core pseudo-potential (DSPP), has been used [16]. These norm-conserving pseudo-potentials are generated by fitting all-electron relativistic DFT results and have a non-local contribution for each channel up to $l$ = 2, as well as a non-local contribution to account for higher channels. To simulate periodic boundary conditions, a vacuum layer of 30 Å was added to the unit cell of the layers. The k-point sampling was done by the use of Monkhorst-Pack scheme [18]. The maximum number of numerical integration mesh points available in DMol3 has been chosen for our computations, and the threshold of density matrix convergence is set to $10^{-6}$.

The existence of magnetic moments in Pu metal is a subject of great controversy and there is no clear compelling evidence, specifically experimental, of magnetic moments in the δ-phase, either ordered or disordered [19]. We also note that, as the films get thicker, the complexity of magnetic ordering, if existent, increases and such calculations can be quite challenging computationally. Nevertheless, to study the effects of spin polarization on the chemisorption process, our studies have been performed at both the spin-polarized and at the non-spin-polarized levels. As for the effects of relativity are concerned, DMol3 does not yet allow fully relativistic computations and as such, we have used the scalar-relativistic approach, as available in Dmol3. In this approach, the effects of spin-orbit coupling is omitted primarily for computational reasons but all other relativistic kinematic effects such as mass-velocity, Darwin, and higher order terms are retained. It has been shown [16] that this approach models actinide bond lengths fairly well. We certainly do not expect that the inclusion of the effects of spin-orbit coupling, though desirable, will alter the primary qualitative and quantitative conclusions of this paper, particularly since we are interested in chemisorption energies defined as the difference in total energies. We also note that Landa *et al.* [20] and Kollar *et al.* [21] have observed that inclusions of spin-orbit coupling are not essential for the quantitative behavior of δ - Pu. Hay and Martin [22] found that one could adequately describe the electronic and geometric properties of actinide complexes without treating spin-orbit effects explicitly. Similar conclusions have been reached by us in our study of water adsorption [12] and of molecular $PuO_2$ and $PuN_2$ [23] and by Ismail *et al*. [24] in their study of uranyl and plutonyl ions. We also note that scalar-relativistic hybrid density functional theory has been used by Kudin *et al*. [25] to describe the insulating gap of $UO_2$, yielding a correct anti-ferromagnetic insulator. All calculations are done on a Compaq ES40 alpha multi-processor supercomputer at the University of Texas at Arlington.

To study oxygen adsorption on Pu surface, the fcc (100) and (111) surfaces are modeled with three layers of Pu at the experimental lattice constant. This is believed to be quite adequate considering that the adatom is not expected to interact with atoms beyond the first three layers. Recently, in a study of quantum size effects in (111) layers of δ- Pu, Ray and Boettger [26] have shown that surface energies converge within the first three layers. Due to severe demands on computational resources, the unit cell per layer is assumed to consist of two Pu atoms. Thus, our three-layer model of the surface contains six Pu atoms. The oxygen atom, one per unit cell, was allowed to approach the Pu surface along four different symmetrical approaches (Figure 1): i) directly on top of a Pu atom (*top* position); ii) on the middle of two nearest neighbor Pu atoms (*bridge* position); iii) in the center of the smallest unit structures of the surfaces (*center* position); and iv) inside the Pu layers (*interstitial* position). The chemisorption energy is calculated from:

$$E_c = E\ (\text{Pu-layers}) + E(\text{O}) - E\ (\text{Pu-layers+O}) \quad (1).$$

For the non-spin-polarized case, both E (Pu-layers) and E (Pu-layers + O) were calculated without spin polarization, while for spin polarized calculations, both of these two energies are spin polarized. E(O) is the energy of the oxygen atom in the ground state. The chemisorption energies, and the corresponding distances are given in table 1.

We first comment on the oxygen adsorption on the $\delta$-Pu (100) surface in the square symmetry without spin polarization. The chemisorption energies as a function of the separation distance of the O atom from the top layer are shown in figures 2(a, c, e, g). We list in table 1 the chemisorption energies and the equilibrium distances of the O atom from the top layer. Clearly, the center site is the most favorable chemisorbed site with chemisorption energy of 7.386eV, followed by the bridge, top and the interstitial sites, with chemisorption energies of 7.065eV, 6.470eV, and 5.422eV respectively. For the center position, the distance of the oxygen atom from the top layer has the lowest value of 0.92 Å, with four Pu atoms at the corners of the square being 2.332 Å apart. For the bridge position, the distance of the oxygen atom from the surface is 1.41 Å, and the nearest O-Pu distance is 2.069 Å. For the top position, where oxygen atom is directly on top of one of the Pu atom, the distance is 1.83 Å. In view of the above picture of distance verses chemisorption energy, we conclude that the most favorable site for chemisorption is determined by the coordination of the oxygen atom with the Pu atoms. For the center site, the coordination number is four to be compared with the coordination number of one for the top site. The situation is similar for the spin polarized case (figures 2(b, d, f, h). With the inclusion of the spin polarization, the chemisorption energies are consistently lower than the non-spin polarized case, and also the adsorption distances are slightly higher in spin polarized case. For example for the most favorable site, center site, chemisorption energies with and without spin polarization are 7.080 eV and 7.386 eV, while the distances of oxygen atom from the plutonium surface is 1.02 Å and 0.92 Å, respectively. For the top site, adsorption without spin polarization is 1.788 eV higher in energy.

Mulliken population analysis [27] indicates (table 2(a)) that for the center position the oxygen atom gains more negative charge compared to the other sites. This indicates that the ionic nature of the bond is stronger for the center site. We also note that for the top site, the Pu atom directly below the oxygen is negatively charged and the surrounding Pu atoms are positively charged. The second layer atoms are more positively charged, comparable to the magnitude of the charge on the oxygen atom. So, for the top site, there is a stronger Coulomb interaction between the oxygen and the second layer Pu atoms. For center and bridge positions, charge distribution is almost similar where the first two layers are positively charged. For the interstitial position, the center of the unit cell is found to be the most favorable position for both spin polarized and non-spin polarized cases; at 2.14 Å below the top face centered atom. The chemisorption energy of this site is 5.422eV (4.963eV with spin polarization), lower than the center and bridge positions. For this site six Pu atoms are at equal distances of 2.14 A from the O atom. Four Pu atoms on the second layer surrounding the O atom are positively charged, whereas the top and below Pu atoms are slightly negatively charged, namely –0.033*e* for non-spin-polarized case. With spin polarization, Pu atoms directly above and below the interstitial oxygen atom are neutral and the interaction is primarily with the other surrounding Pu atoms.

We next consider the (111) surface of fcc δ-Pu. The smallest unit of this surface is an equilateral triangle, so here again the top, bridge, center, and the interstitial sites are the symmetrically distinguishable sites. We first comment on the non-spin-polarized case. Figure 1 shows the different chemisorption sites and figures 3(a, c, e, g) shows the variation of chemisorption energies with the adatom distances to the surfaces. Here the center position is the center of the triangle, which is also the most favorable site with oxygen chemisorption energy of 7.070eV and a distance from the surface of 1.31 Å. This is followed by the bridge site with chemisorption energy of 6.856eV and the distance to the surface is 1.44 Å. For the top site, the chemisorption energy is 6.160eV and the corresponding distance is 1.83 Å. We note that for the (111) surface, the chemisorption energies are consistently lower compared to the energies for the (100) surface, part of the reason being attributed to the fact of different coordination numbers. In (111) surface, the atoms are denser than the (100) surface. This, in conjunction with the fact that chemisorption energies are always lower than those of the (100) surface and the larger Pu-O bond length became the most favorable site, indicate the fact that as oxygen atom comes nearer to the Pu atom, some anti-bonding may play a role. The interstitial site was found directly 1.22 Å below the center of the equilateral triangle with a chemisorption energy of 5.334eV. We also observe that for the chemisorption sites discussed so far for both (100) and (111) surfaces, except for positions of the adatom below the surface, the following inequality between the adatom distance from the surface (r) and corresponding chemisorption energy (C.E.) holds true:

$$r\,(\text{center}) < r(\text{bridge}) < r(\text{top})$$

$$\text{C.E.}(\text{center}) > \text{C.E.}(\text{bridge}) > \text{C.E.}(\text{top}) \quad (2)$$

This implies, as is to be expected, that the highest chemisorption energy is obtained when the adatom is nearest to the surface.

Spin-polarized calculations of (111) surface (figures 3(b, d, f, h)) predict bridge site is slightly more favorable than the center site, as opposed to the non-spin polarized calculations of (111) surface and violates the inequality in equation (2), i.e., distance of oxygen atom from the surface is 0.13 Å higher for bridge site than the center site. However, the optimum chemisorption distances are very close to the corresponding values for the non-spin-polarized sites. Also unlike all other sites, for the bridge, center and the interstitial sites, the spin-polarized chemisorption energies are higher than the corresponding non-spin-polarized energies. The differences between the non-spin-polarized and spin-polarized cases in (111) surface, unlike (100) surface, arises from the fact the, for (111) surface, as mentioned earlier, atoms are denser and the inter-layer separation is lower compared to the (100) surface. Spin plays a stronger role in the (111) surface, as detailed below also and as a result, the effect on the chemisorption process is higher. Population analysis (table 2(b)) also indicates, for the (111) surface, that the charges of the first layers are significantly modified by the presence of the oxygen atom, whereas the effects on the second and third layers are not strong (except for the interstitial cases). Thus, chemisorption activities mainly take place on the first layer with a smaller contribution from the second layer, and the effects decay quickly. Except for the top position, for all the other cases the first two layers are positively charged, and the third layer is negatively charged. This fact is consistent with the fact that for bare Pu layers, the first and the third layers are negatively charged and the middle layer is positively charged. As the adatom approaches, it gains electronic charge at the expense of the first layer, leaving the other two layers only slightly modified. However, for the interstitial cases as expected, charge distributions of all the three layers are affected.

To study the effects of spin-polarization, we have listed in table1 spin magnetic moments for different chemisorption sites of the oxygen adatom. The net magnetic moments of bare plutonium layers are 2.092 $\mu_B$/atom and 1.986 $\mu_B$/atom for the 3-layer (100) and (111) surfaces, respectively. Adsorption of oxygen on plutonium surface reduces the magnetic moments of the system. For (100) surface the magnetic moments increase as the distance of oxygen atom from the surface decreases. For example, for the top site where the oxygen is at highest distance from the Pu surface, the magnetic moment is 1.56 $\mu_B$/atom; whereas for the nearest O-Pu surface distance, the center site, the moment is 1.65 $\mu_B$/atom. The magnetic moment has the highest value for the interstitial position. For the (111) surface, the pattern is not so clear. For the (111) bridge position, the magnetic moment is only 0.12 $\mu_B$/atom, where the top and center sites moments are 1.30 $\mu_B$/atom and 1.38 $\mu_B$/atom, respectively. For the interstitial position of (111) surface the moment drops to almost zero. For both surfaces, spin moments have alternating behavior. For (100) surface, first layer and third layer have up spin, and the second layer has down spin, whereas (111) surface has the opposite behavior. However, though the ordering of spin for (100) surface did not change due to the oxygen adsorption, the ordering of spin was affected for (111) surface. For the center and top positions at first layer both the atoms have parallel spins, but for second and third layer they are anti-parallel. For the bridge and interstitial positions, spins on the plutonium surface have anti-ferromagnetic ordering, which explains the very low net spin magnetic moment in these cases. For all cases, the chemisorbed oxygen atom has very low spin magnetic moment.

A study of the energy levels of the Pu layers before oxygen adsorption indicates that while the 12 6s and 36 6p electrons are localized, the 36 5f electrons appear to be delocalized. The degree of localization decreases as one approaches the Fermi level, indicating the nature of the 5f electrons, in regards to localization versus delocalization depends critically on the electronic positions. Around the Fermi level, the 5f electrons are largely delocalized. This contradicts earlier assertions [7] that Pu 5f electrons are well localized. We do find hybridization of the Pu 7s electrons with the 6d electrons, indicating that the Pu valence behavior might be dominated by the 6d electrons, in agreement with Eriksson *et al* [7]. From the band energetics of the non-spin-polarized bare (100) Pu layers, the top of the 5*f* band is found to be 0.224eV below the Fermi level, and the corresponding value for the (111) surface is 0.328eV. The Fermi level is basically formed by the 7*s* orbitals. We found for (100) surface that the energy gap between 6*s* and 6*p* band is 24.499eV, and that of 6*p* and 5*f* band is 15.404eV. For (111) surface these gaps are 24.581eV and 15.500eV respectively. Upon O adsorption in the center position, the O 1s electron levels are deep core levels, followed by the Pu 6s electrons, the width of the 6s band changing slightly to 0.292eV. However, the adsorption of oxygen atom increases the energy difference between the 5*f* band and the Fermi energy. For the (100) center position, which is the most favorable chemisorption site in this study, this difference is increased to 0.241eV; and for (111) center position the difference is 0.368eV. Similar conclusion is true for the spin-polarized (100) surface. For the center site, after oxygen adsorption the difference between the top of 5f orbital and the Fermi energy is 0.448eV; whereas for the (100) bare Pu layers the difference is 0.391eV. Spin-polarized (111) surface gives a different picture. For example, for the bridge and the center position of oxygen adsorbed (111) surface, the difference between the Fermi energy and 5f orbital is 0.265eV and 0.372eV, respectively, whereas for bare (111) plutonium surface 5f orbital is 0.417eV below the Fermi energy. In this case, 5f orbitals are closer to the Fermi surface due to oxygen adsorptions. Also from the modification of the band energetic due to the presence of the oxygen, it appears that the bonding between the oxygen and the Pu atom is due to the hybridization of Pu 5f and O 2p orbitals, as in our previous cluster study of PuO [12]. Again, the degree of delocalization of the 5f electrons increases as they approach the Fermi level.

In table 3, we tabulate the change in work function due to the oxygen adsorption on the Pu surfaces. It was found that oxygen chemisorbed above the Pu surface has higher work function than the pure Pu surfaces and interstitial position has lower work function, which contradicts earlier assertions in the literature [7]. The change in surface dipole moment due to the presence of oxygen, as is evident from the Mulliken charge distribution, changes the work function of the surface. The (100) center position has minimum increase in the work function. For (111) surface the work function change in bridge and center positions are comparable. In general work function change in spin polarized case is higher than the corresponding non-spin polarized case.

In conclusion, we have studied oxygen adsorption on $\delta$-Pu (100) and (111) surfaces using generalized gradient approximation to density functional theory with Perdew and Wang functionals. The center position of the (100) surface is found to be the most favorable site with the oxygen atom closest to the surface for both spin-polarized and non-spin-polarized cases. Also, among all the cases studied here, non-spin-polarized calculations of the center site gives the highest chemisorption energy. For the (111) surface, the center position is also the preferred site for non-spin-polarized calculations, but for spin polarized calculations, the bridge and the center sites are basically degenerate. It was inferred that 5*f* orbitals are delocalized, specifically as one approaches the Fermi level. Also due to the adsorption of oxygen on both spin polarized and non-spin polarized (100) surface and non-spin polarized (111) surface, plutonium 5f orbital is pushed further below the Fermi energy, compared to the bare plutonium layers; while the spin polarized (111) surface showed the opposite behavior. For (111) surface the effect of spin polarization is more prominent, possibly due to the increased overlap of spinors compared to the (100) surface. The coordination numbers are found to have a significant role in the chemical bonding process. Mulliken charge distribution analysis indicates that the interaction of Pu with O mainly takes place in the first layer with the other two layers being only slightly affected. Work functions, in general, tend to increase due to the presence of oxygen adatom.

This work is supported by the Chemical Sciences, Geosciences and Biosciences Division, Office of Basic Energy Sciences, Office of Science, U. S. Department of Energy (Grant No. DE-FG02-03ER15409) and the Welch Foundation, Houston, Texas (Grant No. Y-1525).

Table 1. Oxygen chemisorption energies (in eV) and distances (in Å) from the Pu surfaces for different positions.

| Surface | Sites | Spin non-polarization | | Spin-Polarized | | Magnetic Moment (in $\mu_B$/atom) |
|---|---|---|---|---|---|---|
| | | Chemisorption Energy (eV) | Distances (Å) | Chemisorption Energy (eV) | Distances (Å) | |
| (100) | Top | 6.470 | 1.83 | 4.682 | 1.85 | 1.56 |
| | Bridge | 7.065 | 1.41 | 6.700 | 1.45 | 1.58 |
| | Center | 7.386 | 0.92 | 7.080 | 1.02 | 1.65 |
| | Interstitial | 5.422 | 2.14 | 4.936 | 2.14 | 1.92 |
| (111) | Top | 6.160 | 1.83 | 6.140 | 1.84 | 1.30 |
| | Bridge | 6.856 | 1.44 | 7.238 | 1.46 | 0.12 |
| | Center | 7.070 | 1.31 | 7.217 | 1.33 | 1.38 |
| | Interstitial | 5.334 | 1.22 | 5.510 | 1.23 | 0.01 |

Table 2(a). Mulliken charge distributions for different chemisorption sites for (100) surface. The first column of number are the charge distribution of the Pu three layers without oxygen atom. The other four columns are the different chemisorption sites. NSP indicate no spin polarization and SP is with spin polarization.

| | Layers | without O | Top | Bridge | Center | Interstitial |
|---|---|---|---|---|---|---|
| NSP | O-atom | × | -0.337 | -0.571 | -0.610 | -0.529 |
| | 1st layer | -0.115 | -0.100 | 0.172 | 0.243 | -0.176 |
| | | -0.115 | 0.220 | 0.172 | 0.221 | -0.033 |
| | 2nd layer | 0.230 | 0.206 | 0.218 | 0.227 | 0.806 |
| | | 0.230 | 0.261 | 0.218 | 0.211 | 0.141 |
| | 3rd layer | -0.115 | -0.112 | -0.130 | -0.149 | -0.176 |
| | | -0.115 | -0.138 | -0.130 | -0.142 | -0.033 |
| SP | O-atom | × | -0.384 | -0.572 | -0.665 | -0.610 |
| | 1st layer | -0.094 | -0.055 | 0.224 | 0.221 | -0.174 |
| | | -0.094 | 0.240 | 0.224 | 0.359 | 0.000 |
| | 2nd layer | 0.188 | 0.179 | 0.169 | 0.154 | 0.719 |
| | | 0.188 | 0.224 | 0.169 | 0.186 | 0.239 |
| | 3rd layer | -0.094 | -0.085 | -0.107 | -0.129 | -0.174 |
| | | -0.094 | -0.119 | -0.107 | -0.129 | 0.000 |

Table 2(b). Mulliken charge distributions for different chemisorption sites for (111) surface. The first column of number are the charge distribution of the Pu three layers without oxygen atom. The other three columns are the different chemisorption sites. NSP indicate no spin polarization and SP is with spin polarization.

| | Layers | without O | Top | Bridge | Center | Interstitial |
|---|---|---|---|---|---|---|
| NSP | O-atom | × | -0.305 | -0.510 | -0.600 | -0.516 |
| | $1^{st}$ layer | -0.130 | -0.083 | 0.171 | 0.288 | 0.062 |
| | | -0.130 | 0.122 | 0.161 | 0.131 | -0.051 |
| | $2^{nd}$ layer | 0.260 | 0.241 | 0.234 | 0.263 | 0.497 |
| | | 0.260 | 0.288 | 0.225 | 0.204 | 0.351 |
| | $3^{rd}$ layer | -0.130 | -0.138 | -0.149 | -0.148 | -0.186 |
| | | -0.130 | -0.126 | -0.131 | -0.139 | -0.158 |
| SP | O-atom | × | -0.384 | -0.557 | -0.653 | -0.587 |
| | $1^{st}$ layer | -0.094 | -0.052 | 0.215 | 0.335 | 0.105 |
| | | -0.094 | 0.118 | 0.170 | 0.144 | -0.056 |
| | $2^{nd}$ layer | 0.188 | 0.251 | 0.232 | 0.265 | 0.545 |
| | | 0.188 | 0.284 | 0.219 | 0.201 | 0.349 |
| | $3^{rd}$ layer | -0.094 | -0.136 | -0.153 | -0.147 | -0.186 |
| | | -0.094 | -0.126 | -0.126 | -0.145 | -0.169 |

Table 3. Work function change due to the oxygen chemisorption on Pu surface.

| Sites | Change in work function in eV | | | |
|---|---|---|---|---|
| | (100) surfaces | | (111) surfaces | |
| | NSP | SP | NSP | SP |
| Top | 0.997 | 1.169 | 1.007 | 1.136 |
| Bridge | 0.686 | 0.843 | 0.750 | 0.850 |
| Center | 0.252 | 0.387 | 0.658 | 0.708 |
| Interstitial | −0.326 | −0.254 | −0.179 | −0.202 |

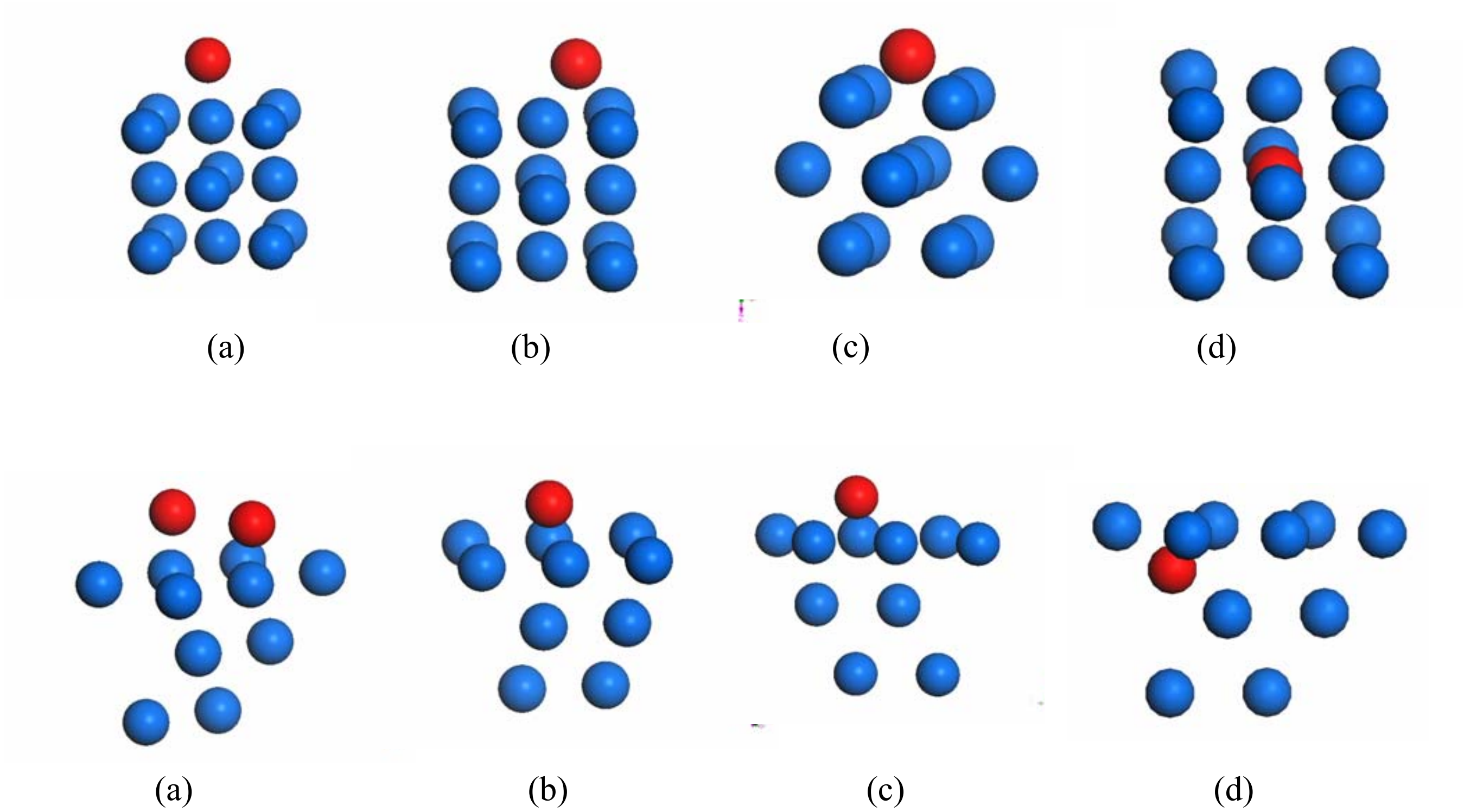

Figure 1. Different chemisorption position for (100) surface in the first row and (111) Surface in the second row: (a) *top*, (b) *bridge,* (c) *center* and (d) *interstitial* positions.

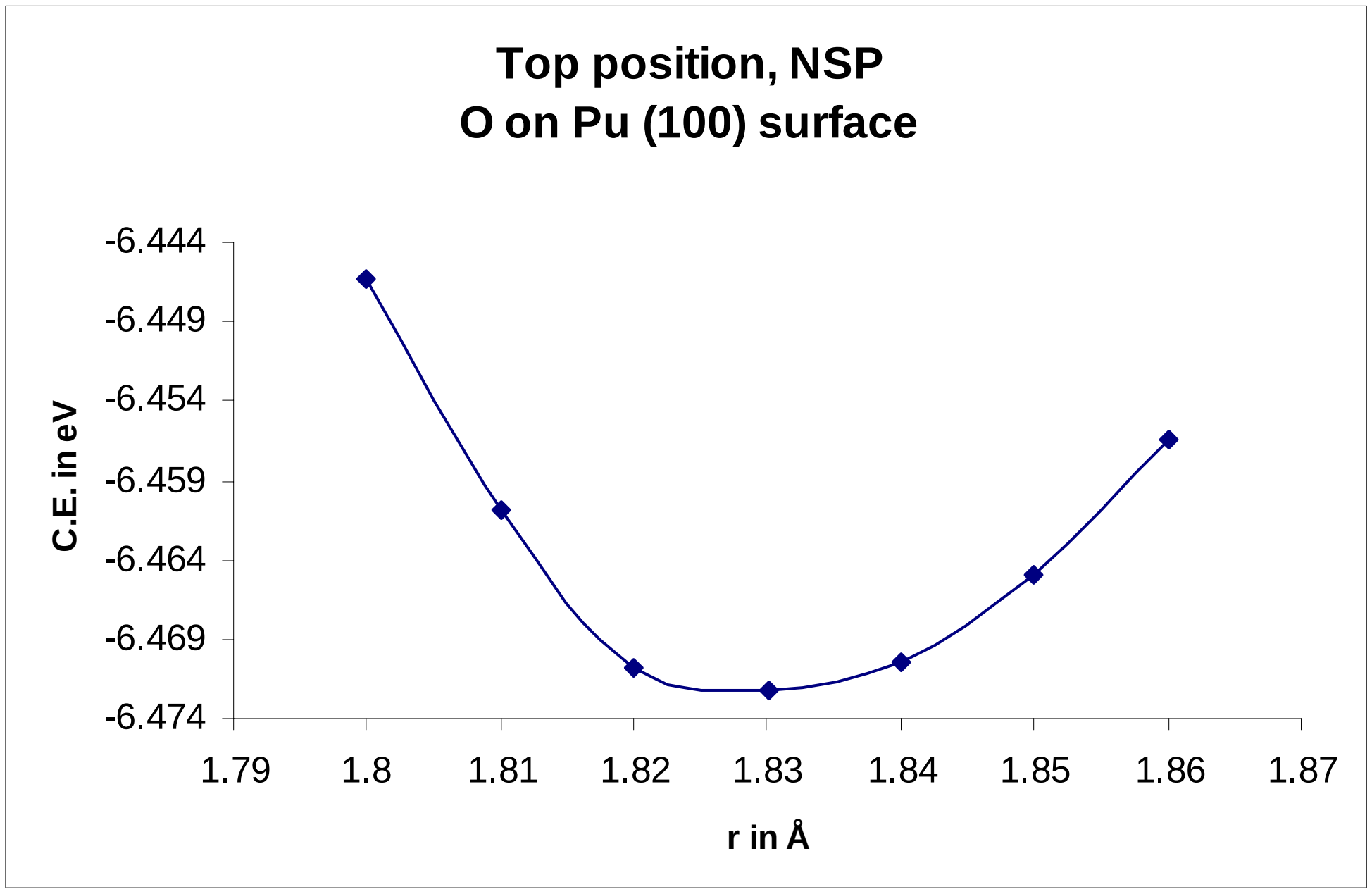

Figure 2(a). Non-spin-polarized chemisorption energy versus the oxygen adatom distance from the Pu (100) surface in the top position.

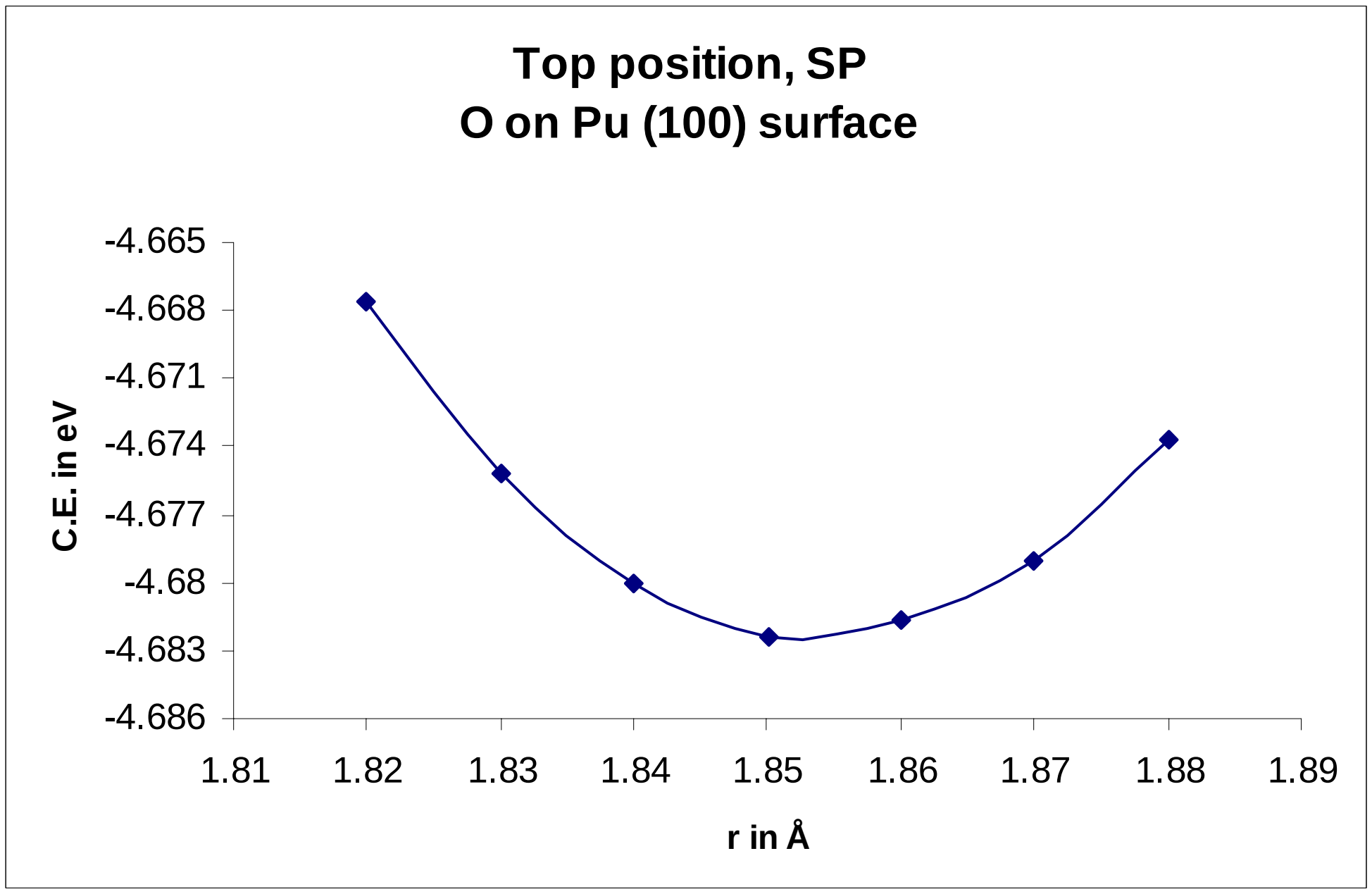

Figure 2(b). Spin-polarized chemisorption energy versus the oxygen adatom distance from the Pu (100) surface in the top position.

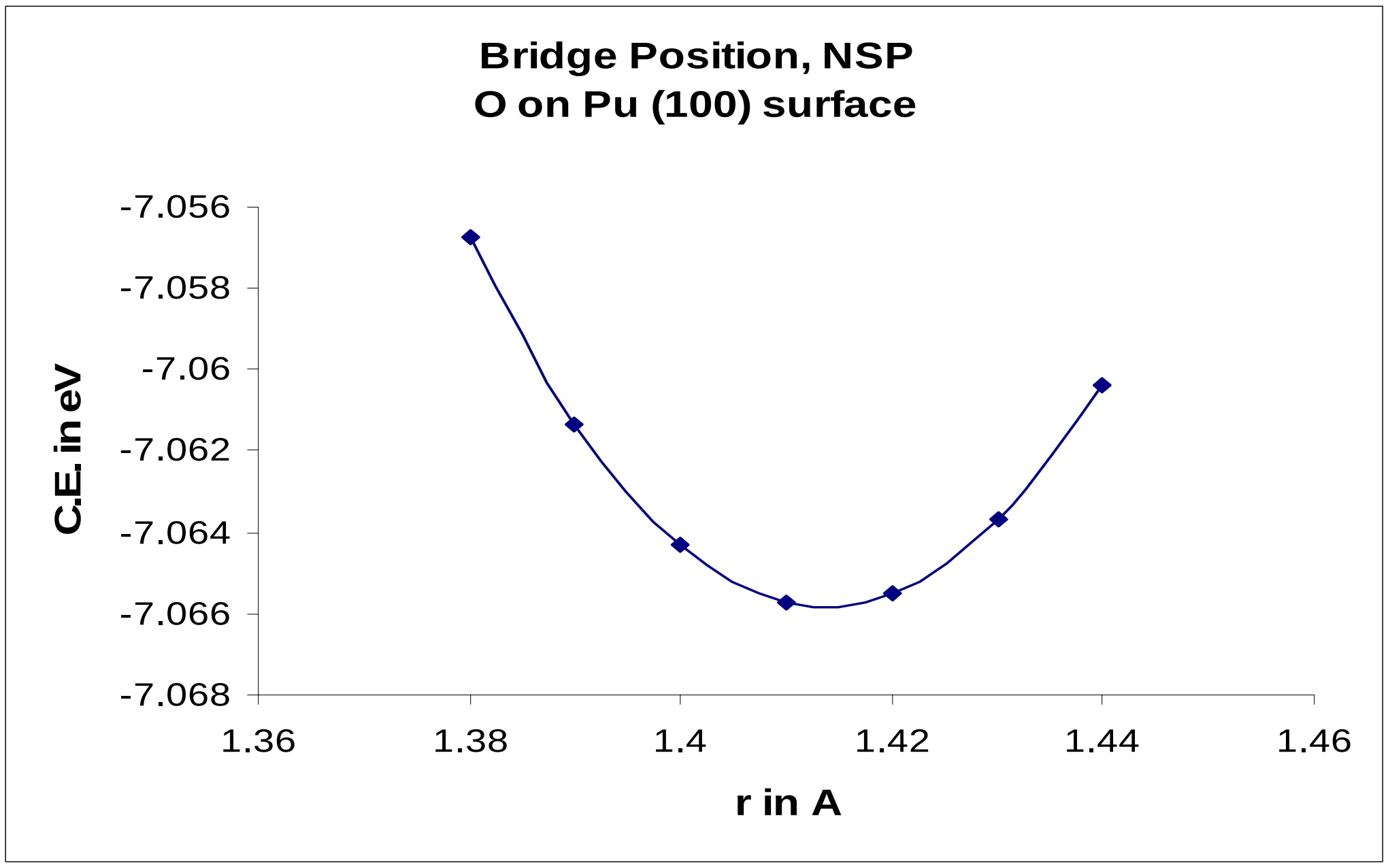

Figure 2(c). Non-spin-polarized chemisorption energy versus the oxygen adatom distance from the Pu (100) surface in the bridge position.

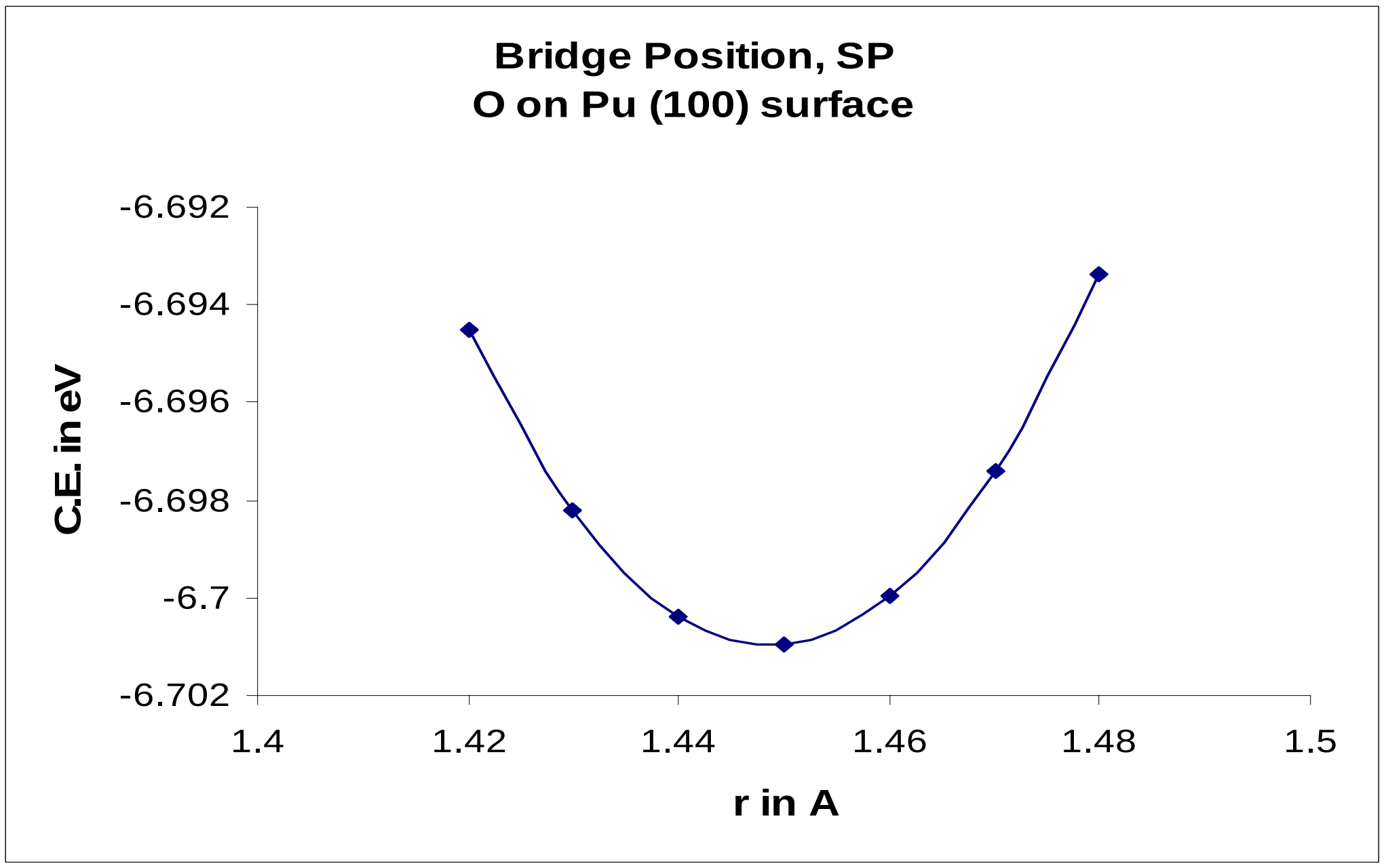

Figure 2(d). Spin-polarized chemisorption energy versus the oxygen adatom distance from the Pu (100) surface in the bridge position.

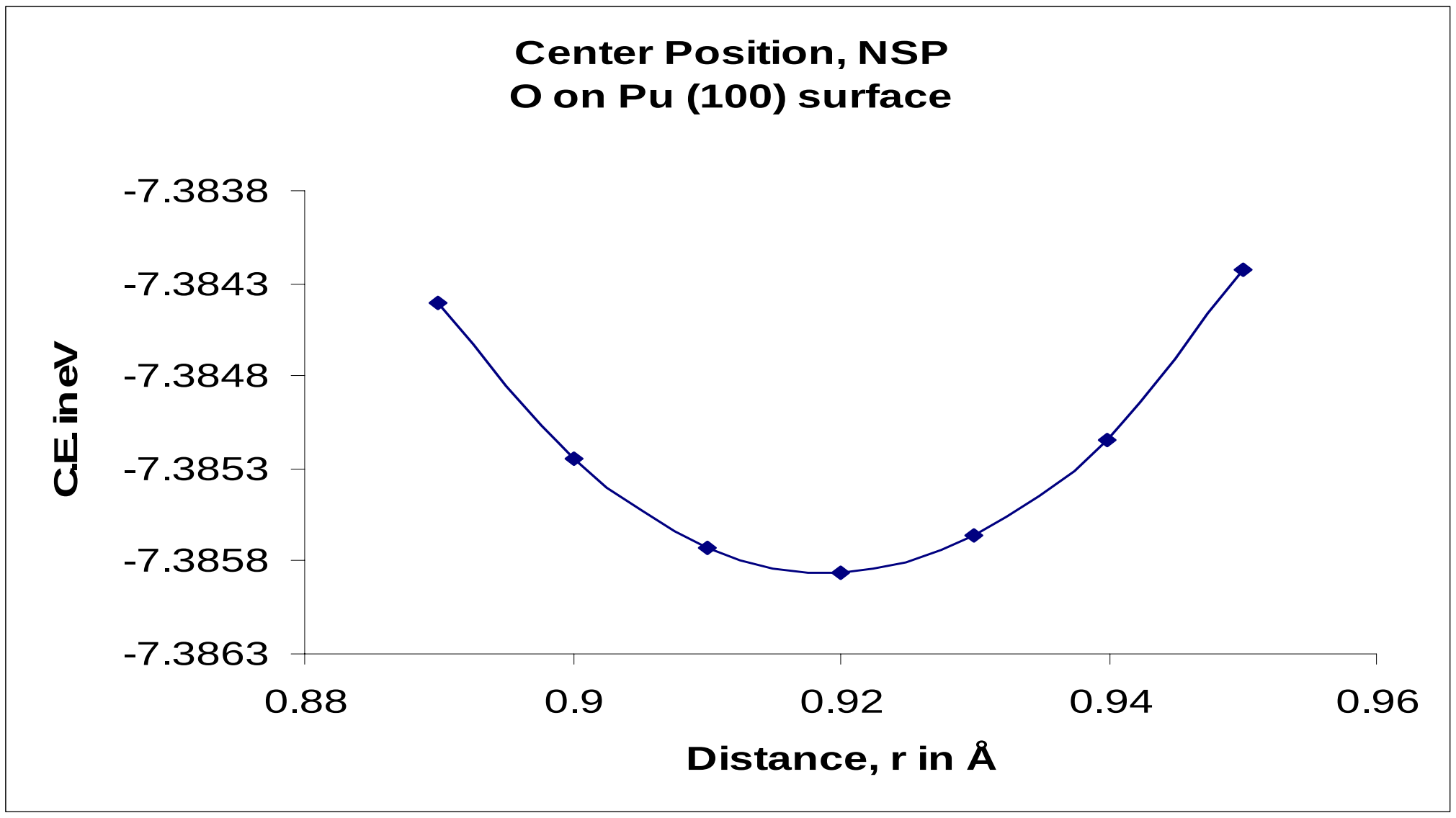

Figure 2(e). Non-spin-polarized chemisorption energy versus the oxygen adatom distance from the Pu (100) surface in the center position.

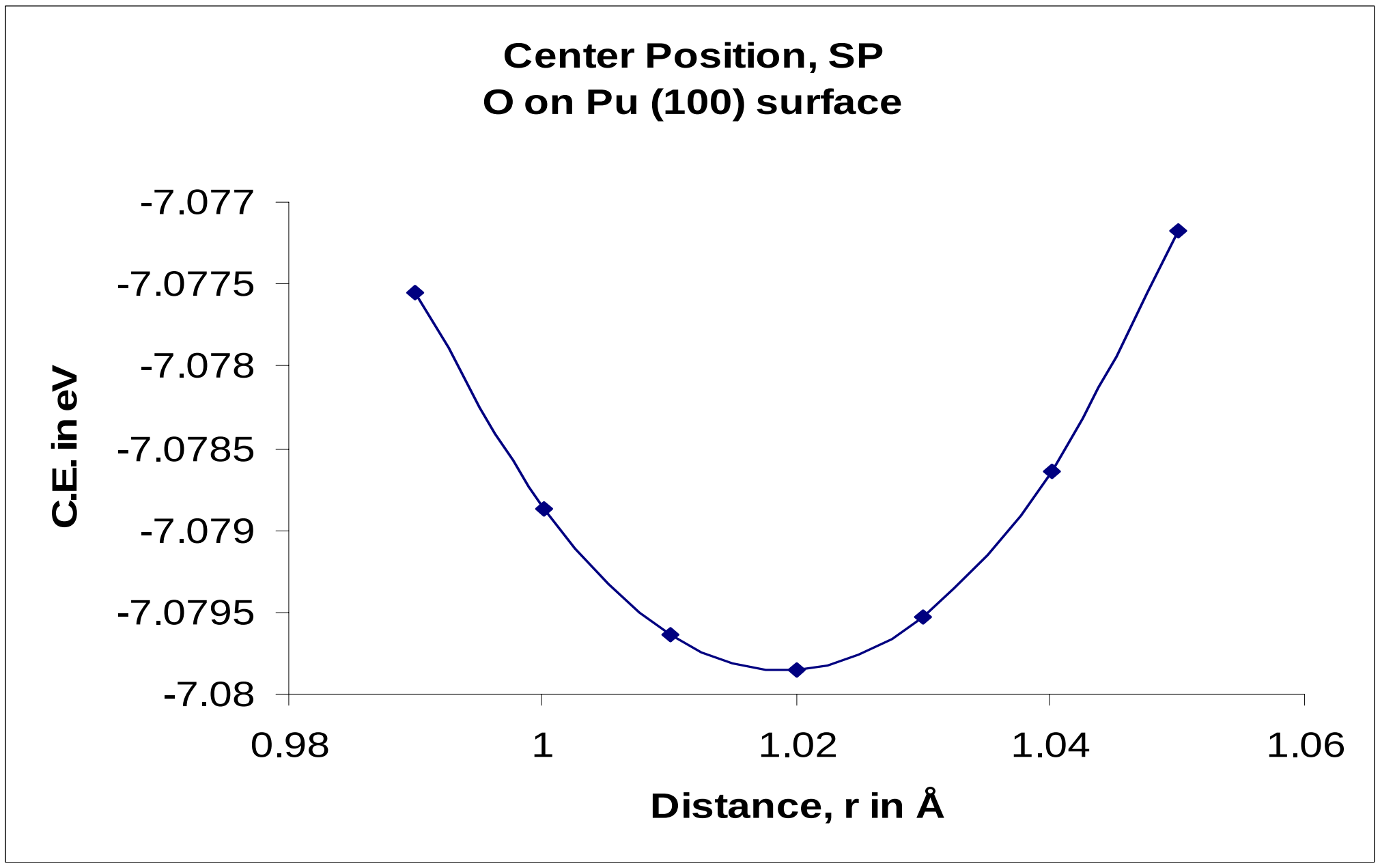

Figure 2(f). Spin-polarized chemisorption energy versus the oxygen adatom distance from the Pu (100) surface in the center position.

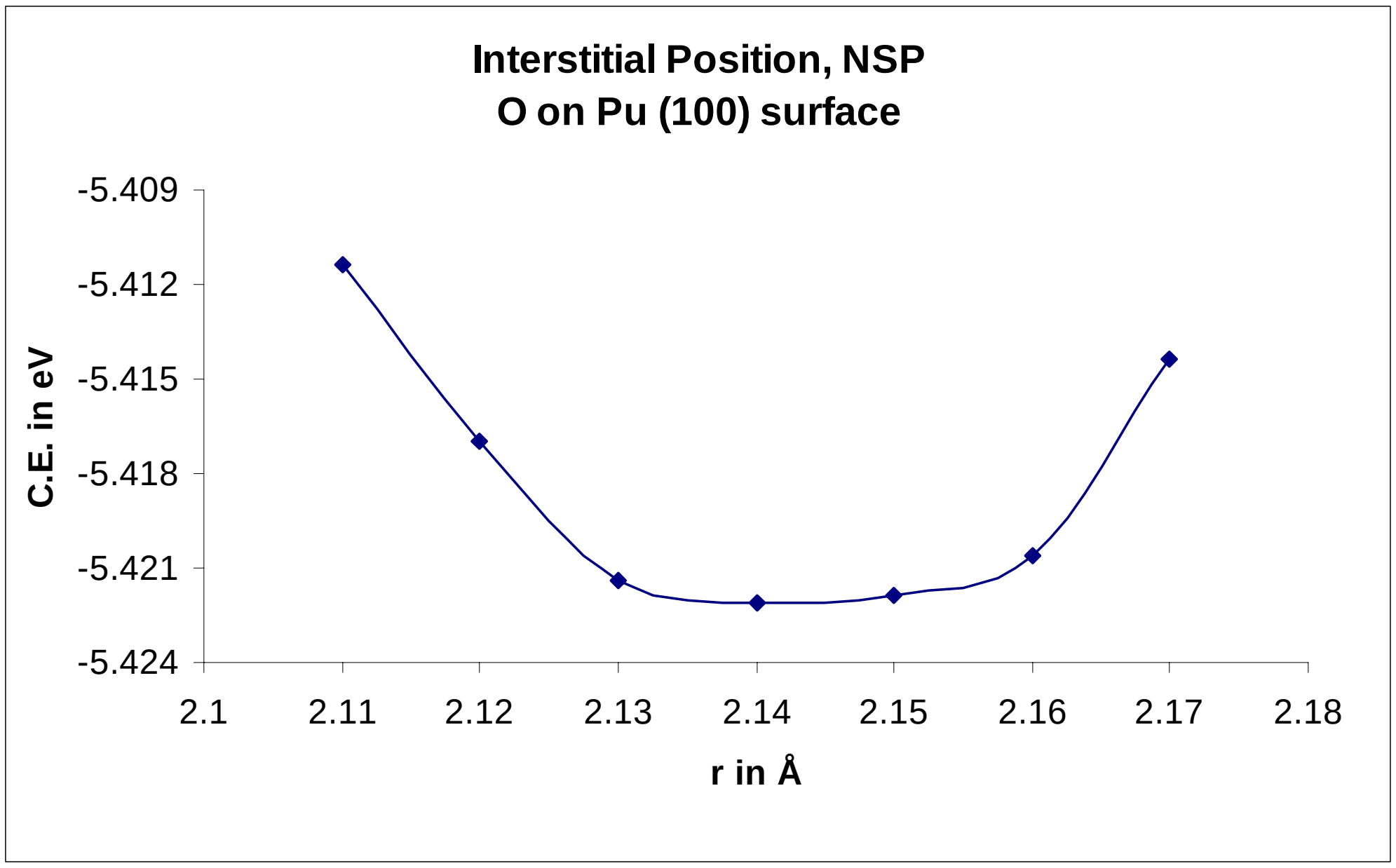

Figure 2(g). Non-spin-polarized chemisorption energy versus the oxygen adatom distance from the Pu (100) surface in the interstitial position.

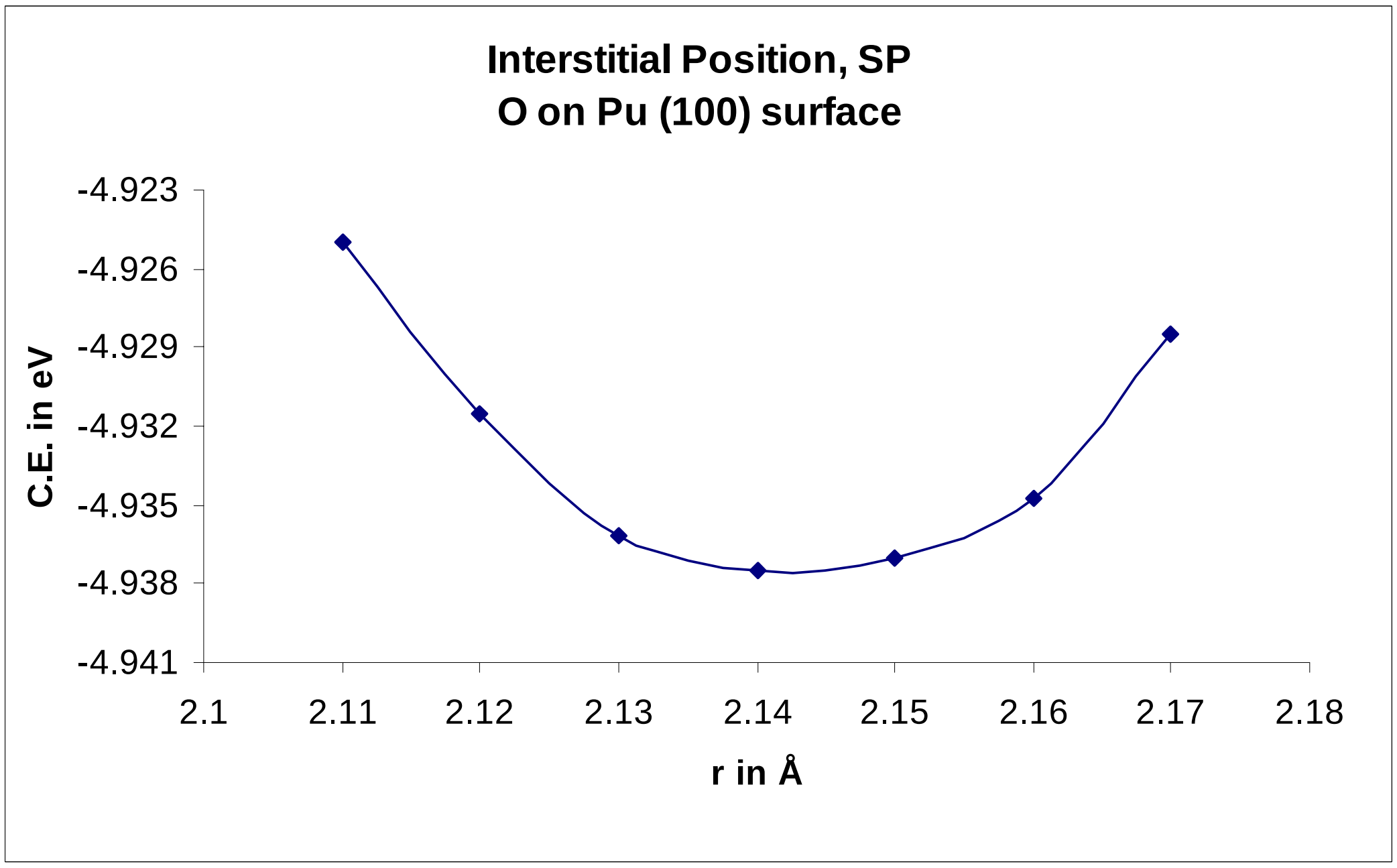

Figure 2(h). Spin-polarized chemisorption energy versus the oxygen adatom distance from the Pu (100) surface in the interstitial position.

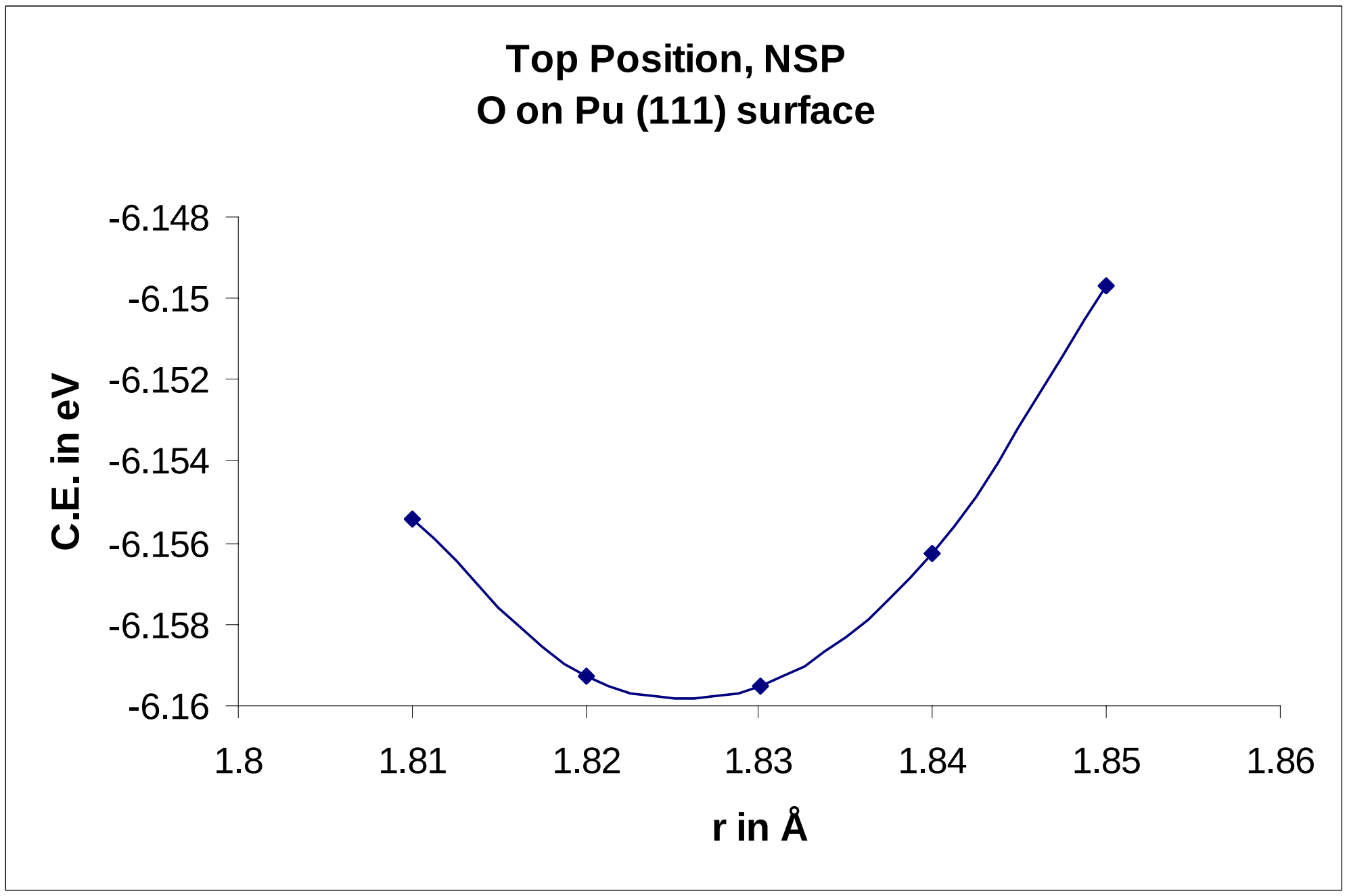

Figure 3(a). Non-spin-polarized chemisorption energy versus the oxygen adatom distance from the Pu (111) surface in the top position.

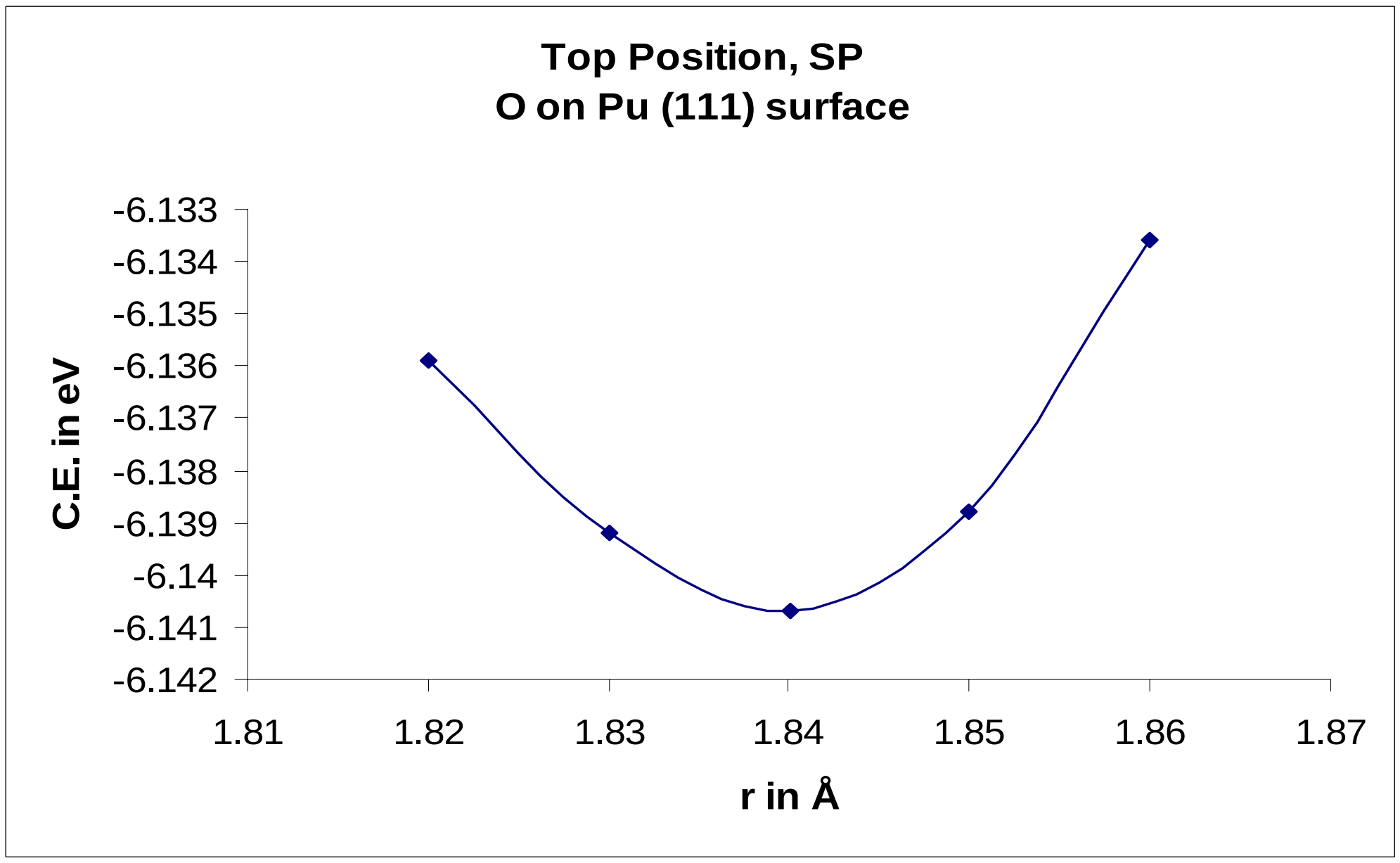

Figure 3(b). Spin-polarized chemisorption energy versus the oxygen adatom distance from the Pu (111) surface in the top position.

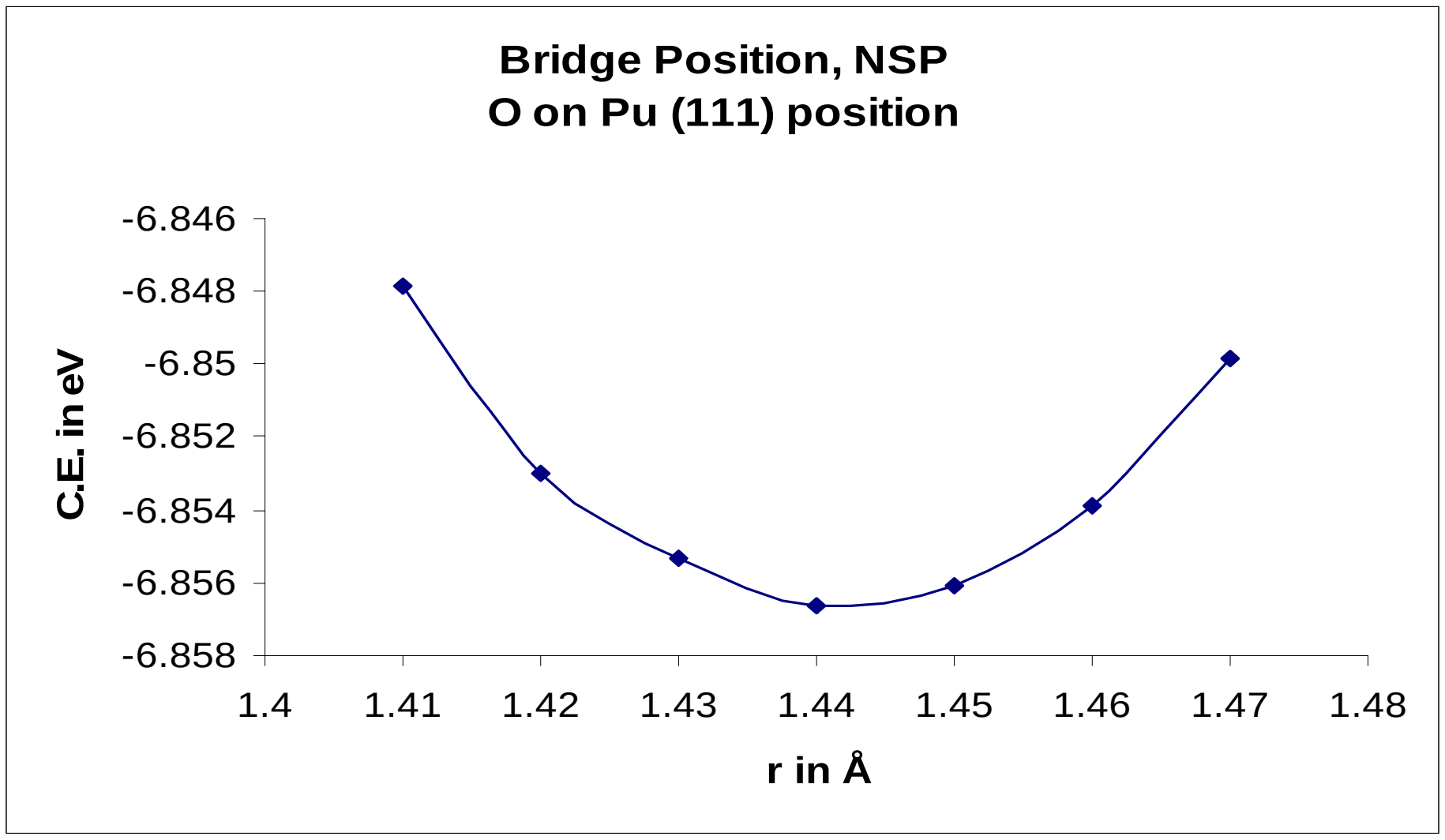

Figure 3(c). Non-spin-polarized chemisorption energy versus the oxygen adatom distance from the Pu (111) surface in the bridge position.

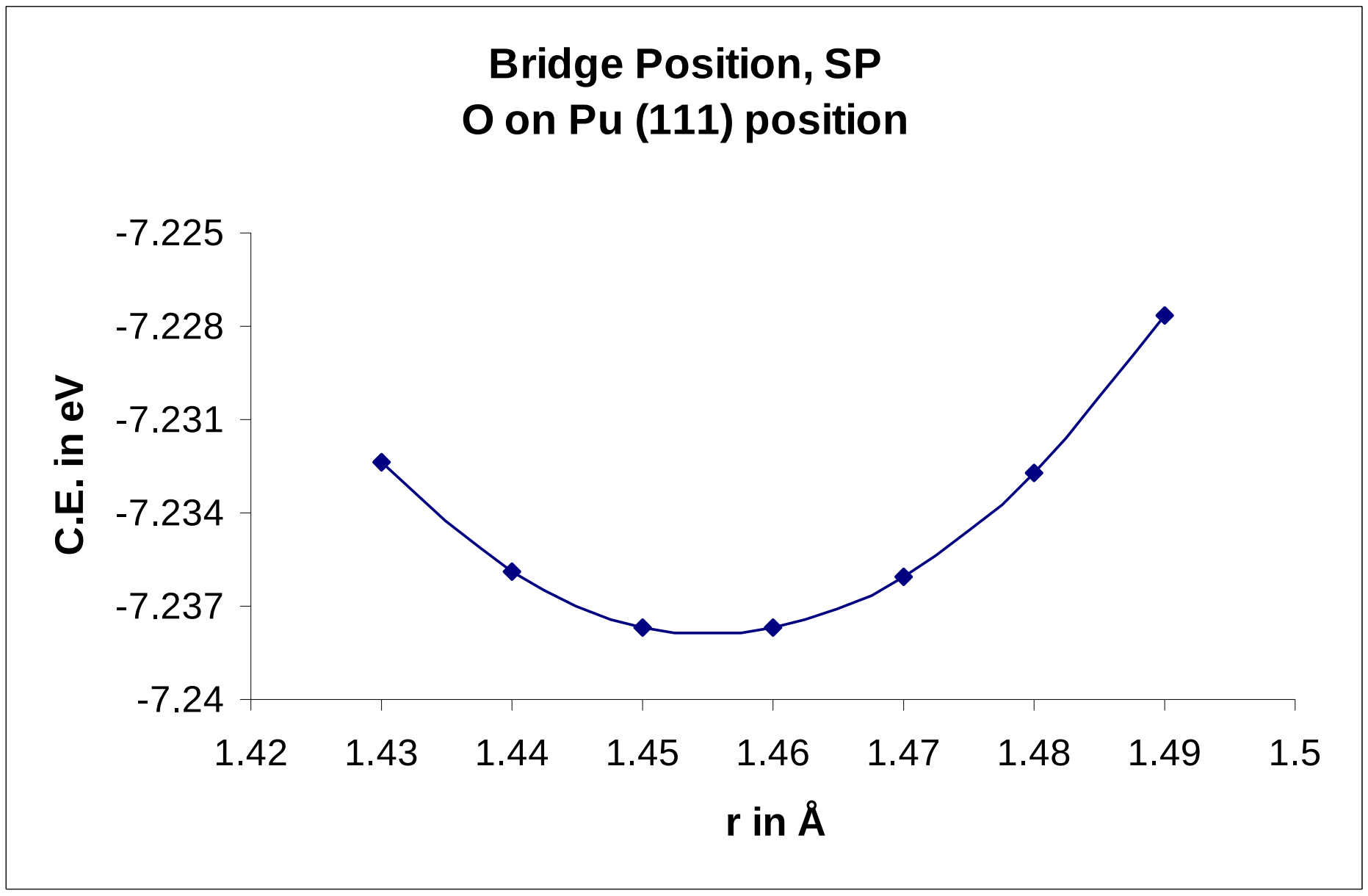

Figure 3(d). Spin-polarized chemisorption energy versus the oxygen adatom distance from the Pu (111) surface in the bridge position.

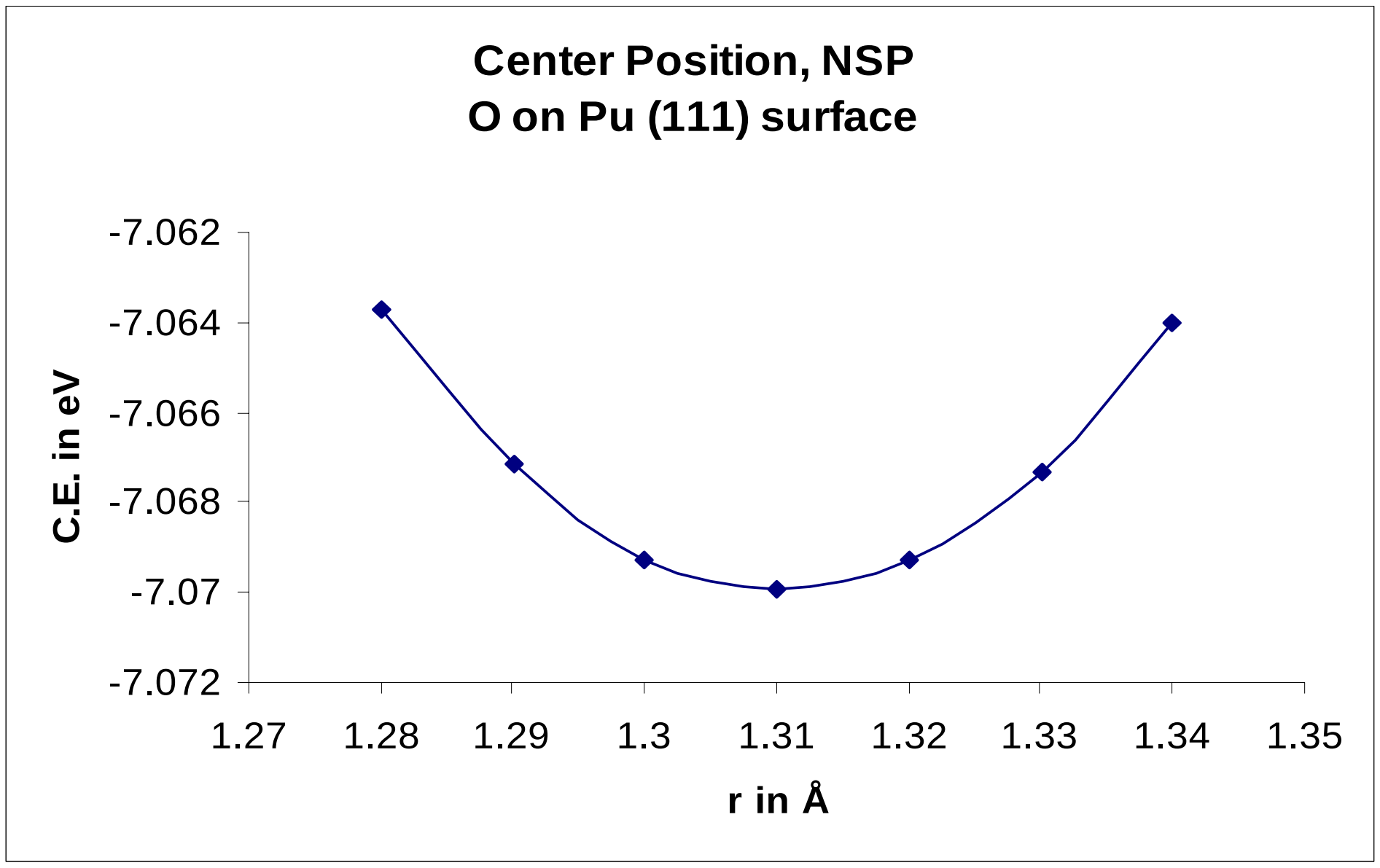

Figure 3(e). Non-spin-polarized chemisorption energy versus the oxygen adatom distance from the Pu (111) surface in the center position.

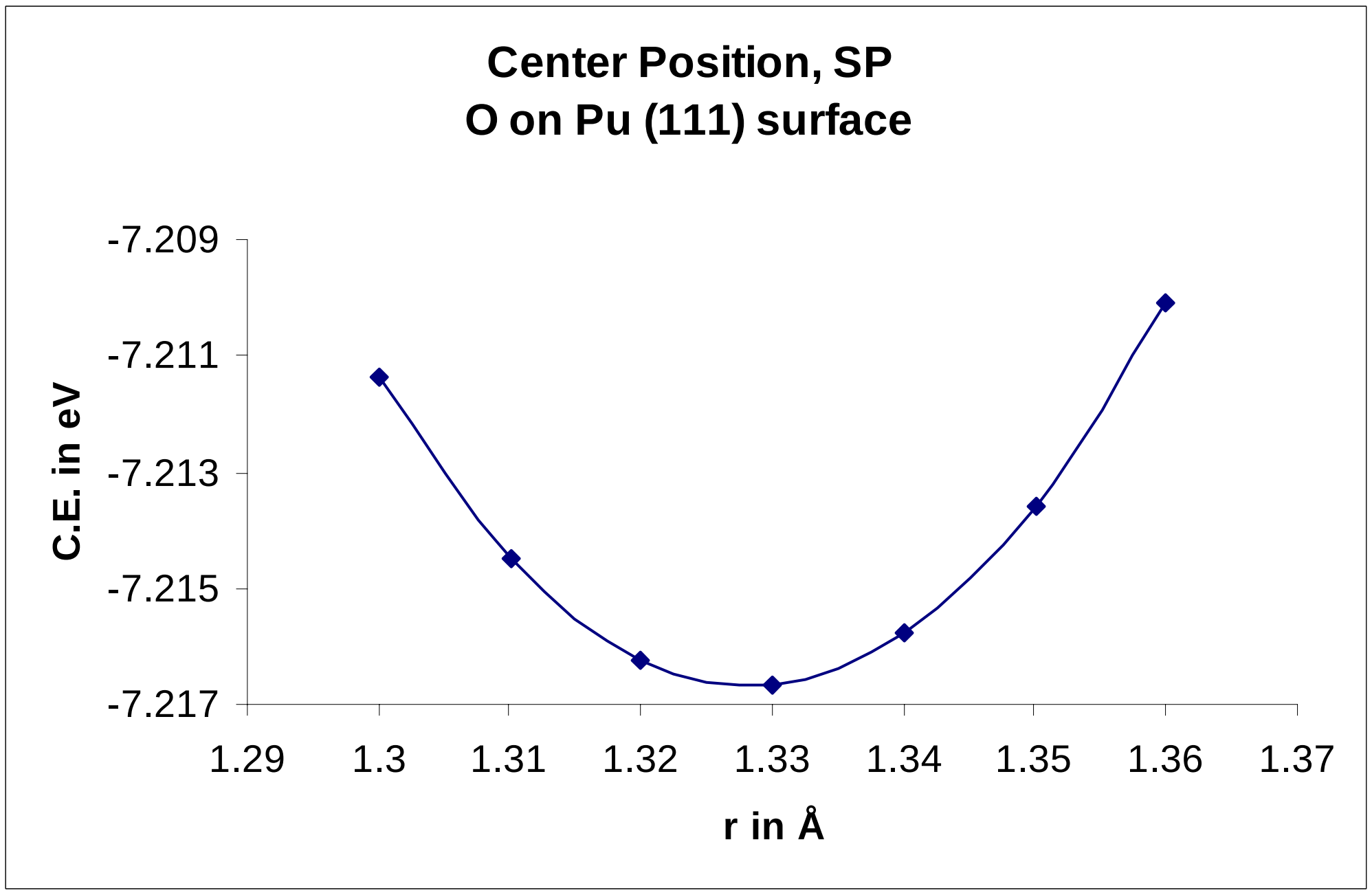

Figure 3(f). Spin-polarized chemisorption energy versus the oxygen adatom distance from the Pu (111) surface in the center position.

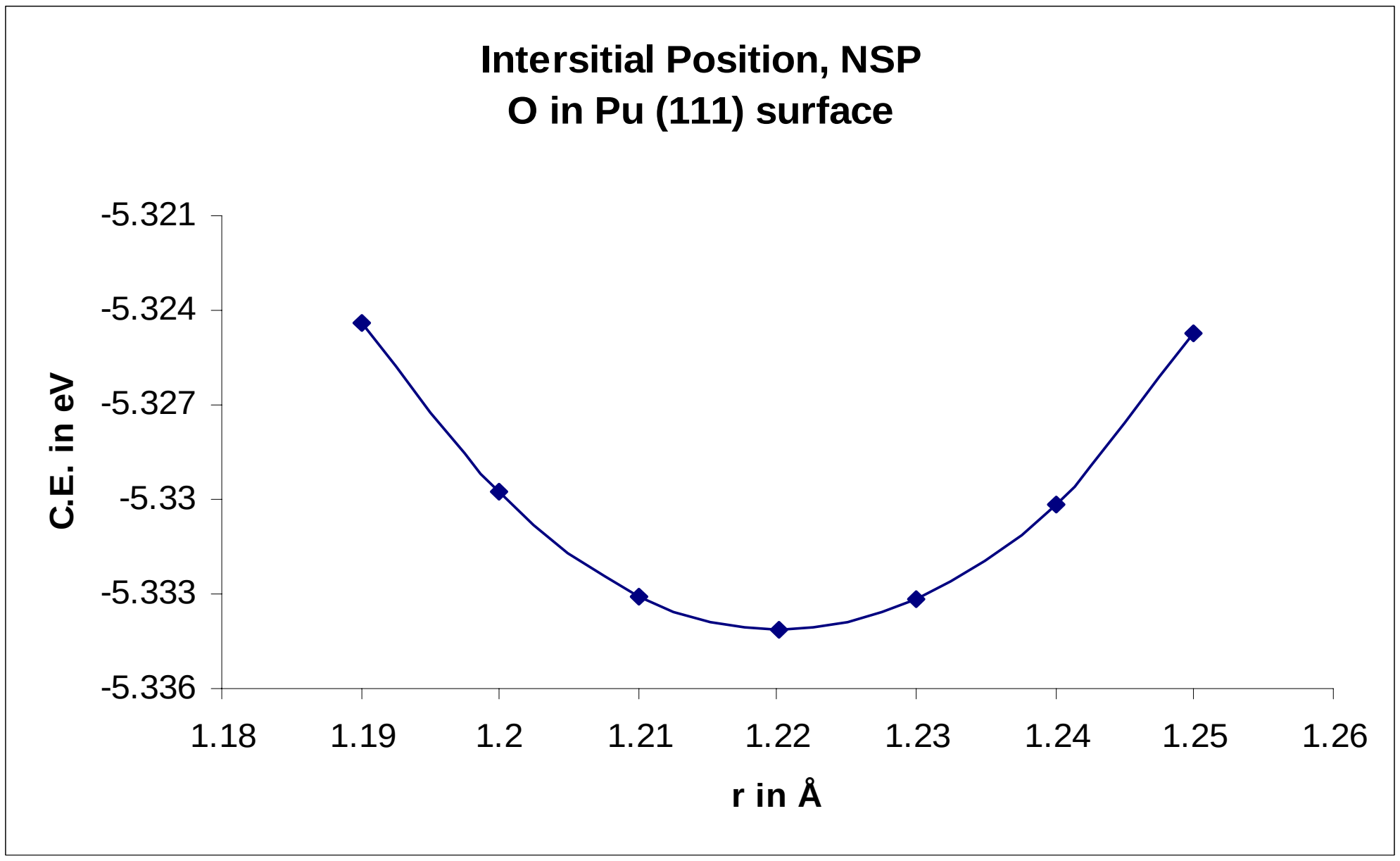

Figure 3(g). Non-spin-polarized chemisorption energy versus the oxygen adatom distance from the Pu (111) surface in the interstitial position.

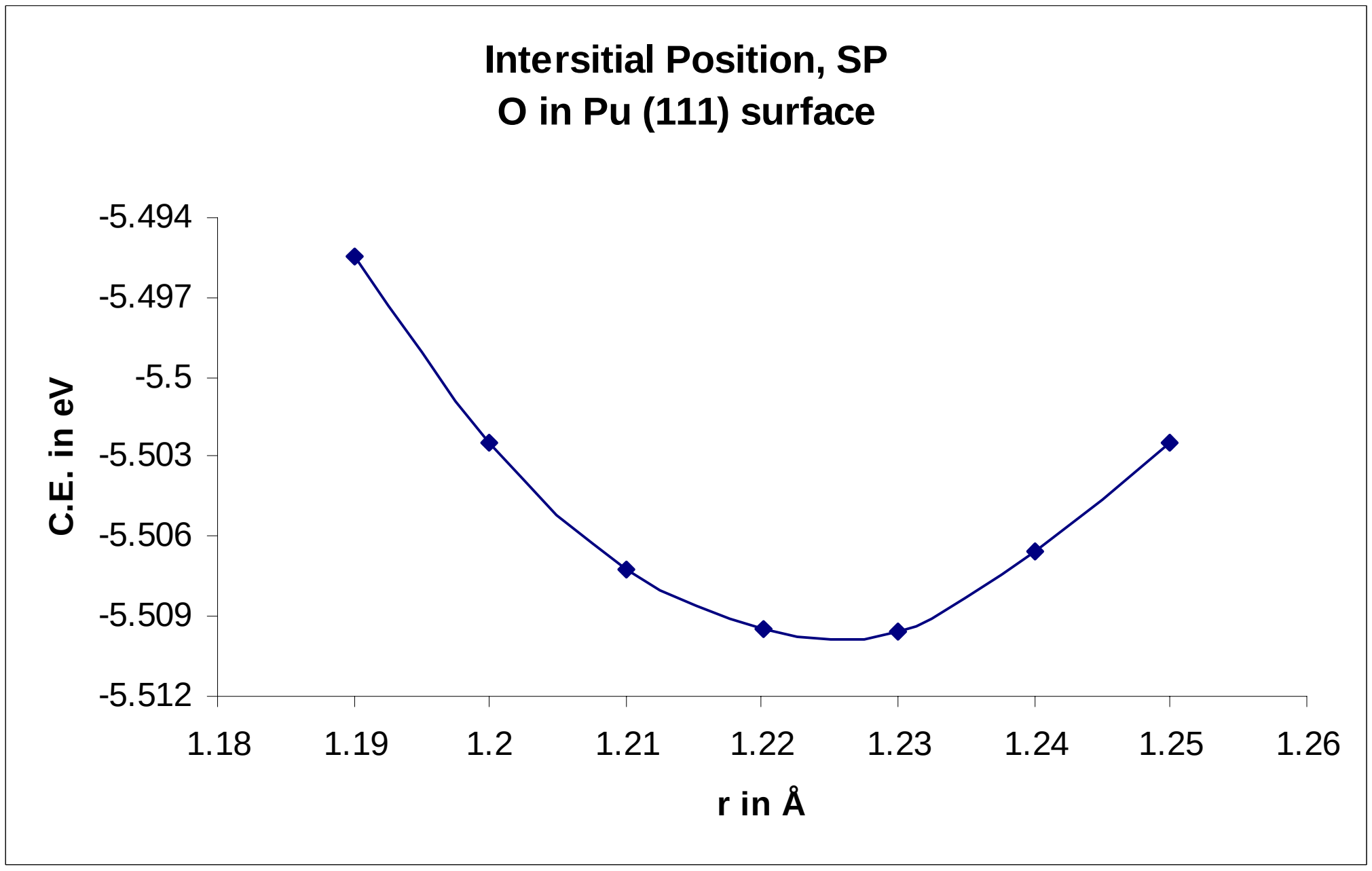

Figure 3(h). Spin-polarized chemisorption energy versus the oxygen adatom distance from the Pu (111) surface in the interstitial position.